# Understanding cooperative loading in carbon nanotube fibres through in-situ structural studies during stretching

Juan C. Fernández-Toribio[a,b], Anastasiia Mikhalchan[b], Cleis Santos[b], Alvaro Ridruejo[a,*], Juan J. Vilatela[b,*]

[a]Department of Materials Science, Universidad Politécnica de Madrid, ETSI Caminos. C/Profesor Aranguren, 3, Madrid, Spain

[b]IMDEA Materials Institute. Eric Kandel, 2, Tecnogetafe, 28906, Getafe, Madrid, Spain

**Abstract**

Carbon nanotube (CNT) fibres are firmly established as a new high-performance fibre, but their tensile mechanical properties remain a relatively small fraction of those of the constituent CNTs. Clear structure-property relations and accurate mechanical models are pressing requirements to bridge this gap. In this work we analyse the structural evolution and molecular stress transfer in CNT fibres by performing in-situ synchrotron wide- and small-angle X-ray scattering and Raman spectroscopy during tensile deformation. The results show that CNT fibres can be accurately described as network of bundles that slide progressively according to the initial orientation distribution function of the material following a Weibull distribution. This model decouples the effects of CNT alignment and degree of cooperative loading, as demonstrated for fibres produced at different draw ratios. It also helps explain the unusually high toughness (fracture energy) of CNT fires produced by the direct spinning method, a key property for impact resistance in structural materials, for example.

*Corresponding author

*Email addresses:*

alvaro.ridruejo@upm.es (Alvaro Ridruejo),

juanjose.vilatela@imdea.org (Juan J. Vilatela)

# 1. Introduction

The development of macroscopic materials based on carbon nanotubes (CNTs) that successfully exploit their superlative properties and rival established engineering materials remains in many cases a challenge. The organization of CNTs into aligned fibres is recognised as a natural embodiment for a 1D material and the most effective to transfer axial properties to the bulk, similar to the architecture of high-performance polymer fibres. Indeed, optimisation of experimental parameters during synthesis and spinning have led to CNT fibres with specific longitudinal tensile strength and modulus in the range of 1.5 – 2.5 GPa/SG and 160 – 250 GPa/SG, respectively, which can further increase after post-pinning chemical treatment [1]. Yet, the continuous improvement in tensile properties of CNT fibres over the last two decades has been achieved almost entirely by trial and error.

Developing a robust micromechanical framework that relates CNT fibre structure to bulk tensile mechanical properties has proven challenging because of the inherently complex hierarchical structure of CNT fibres, and because of large differences in properties of materials produced using difference spinning processes in terms of the constituent CNTs and their degree of alignment. Pioneering work by Espinosa and co-workers has analysed stress transfer by shear between adjacent CNTs and bundles, and proposed simulation models to extrapolate results from in-situ tensile testing of bundles in a transmission electron microscope to the length scale of a CNT fibre (around 10 microns in diameter) [2]. Such work has focused on bridging different length scales of particular CNT fibres, but the role of additional factors controlling fibre modulus and strength impeded widespread comparison and thus more generality.

It was early recognised that stress transfer occurs in shear between CNTs/bundles and that thus tensile properties could benefit from maximizing overlap contact at the nanoscale through more effective self-packing of few-layer CNTs or densification induced through capillary forces or mechanical pressure, and by increasing CNT length [3]. Indeed, a recent study screening tensile properties of CNT fibres spun from liquid crystalline solutions of many different CNTs, has demonstrated that tensile stress strength scales with aspect ratio to the power of 0:9 [4], which is close to an exponent of 1 expected for frictional stress transfer.

Similarly, the degree of alignment of CNTs parallel to the fibre axis has been assumed to be a critical factor controlling bulk properties. In a previous work we demonstrated that CNT fibres can be reduced to a network of bundles and the bulk tensile modulus mathematically related to the orientation distribution function (ODF) of the load-bearing CNT bundles using a the fibrillary crystallite uniform stress transfer model originally developed for rigid-rod polymer fibres and carbon fibre. While on much firmer theoretical footing, this description however, only describes accurately the elastic deformation of CNT fibres [5]. Issues such as their elasto-plastic mechanical and strain-hardening behavior, their extraordinary energy to break (as high as 100$J/g$) and ballistic figure of merit (900$m/s$) and in general a more complete description of their tensile properties are still lacking. Such description is a key tool for the rational application of synthetic and assembly methods in the quest for the next generation multifunctional high-performance fibre.

In this work we use two types of in-situ characterisation methods (synchrotron wide- and small-angle X-ray scattering and Raman spectroscopy) to determine the structural evolution of CNT fibres during tensile testing, and demonstrate that their behaviour can be accurately described as a network of bundles that slide progressively according to the initial ODF of the material following a Weibull distribution.

## 2. Experimental

Carbon nanotube fibres were synthesised by the direct spinning method whereby an aerogel of CNTs is directly drawn out from the gas phase during growth by foating catalyst chemical vapour deposition (CVD) [6]. This chemical reaction takes place at the top part of a vertical tubular furnace at 1250 °C. Butanol was used as carbon source, sulphur as the promoter of the reaction and ferrocene as the catalyst element in a proportion of (97.7:1.5:0.8) adjusted to produced multiwalled CNTs of 3 - 5 layers [7]. The degree of CNT orientation in the fibre was varied by changing the rate at which the fibres were drawn out of the reactor, equivalent to the winding rate [8]. The CNT fibres in this study were extracted from two batches obtained at winding rates of 20 and 30 m/min, respectively. All samples were densified with acetone after spinning and left to dry overnight. *In-situ* mechanical tests were carried out with a modified Kammrath und Weiss miniaturized tensile stage with a load cell with sensitivity of ± 7 nN, at a strain rate of 2 µm/s. Fibre samples of 15 mm were placed onto a rectangular paperboard frame attaching the two end points by a high-viscosity epoxy glue and leaving the rest freestanding. Epoxy infiltration into CNT fibres and curing typically improves their tensile properties [9], nevertheless, we inspected samples subjected to in-situ tests to confirm that fracture occurred remote from the epoxy-infiltrated CNT fibre at the grips. This frame was mechanically clamped to the tensile testing machine. In order to avoid any possible viscoelastic effects, X-ray/Raman measurements were taken after a few minutes of applying strain. Specific stress values were calculated as the force reached by the sample normalized by its linear density, expressed in *N/tex* where *tex = g/km* or, equivalently, *GPa/SG*. The value of linear density was obtained by gravimetric measurements of fibre samples of known length, typically around 30 m. Variation in linear density over 20 mm are of the order of 20%, as determined with a Textechno Favimat using the vibroscopic method.

SAXS and WAXS were carried out at the NCD-SWEET beamline in ALBA synchrotron with radiation wavelength λ=1 Å. Sample-to-detector distance and other parameters were calibrated using reference materials. Data were processed with the software DAWN [10]. Two types of beam configurations were used: macrofocus with spot size at the focal plane of approximately 100 µm × 50 µm and microfocus with diameter of ~ 10 µm. No differences in ODF are found between these modes; the microfocus configuration simply increases scattering intensity over background signal, useful for radial profile analysis. SAXS radial profiles were analysed as a two-phase system (CNTs and air) with sharp boundaries after correction for density fluctuations, as detailed in[11]. Porosity (*P*) was calculated from the total scattering intensity, the invariant (*Q*) being

$$Q = \int_0^\infty q^2 I_{corr}(q) dq = 8\pi P(1-P)(\Delta\rho)^2, \qquad (1)$$

Where *q* is the scattering vector, $I_{corr}(q)$ the scattering intensity after density fluctuation correction, and Δρ the difference in electron density between the two phases.

The CNT orientation distribution ψ(φ) is approximated from the intensity of the azimuthal profile of the SAXS signal after radial integration

$$\Psi(\phi) = \frac{I(\phi)}{\int_0^\pi I(\phi) sin(\phi) d\phi}, \qquad (2)$$

And the orientation parameter <cos²(φ)>

$$< cos^2\phi > = \int_0^\pi cos^2\phi \ \Psi(\phi) \ sin\phi \ d\phi \qquad (3)$$

Strictly speaking, the X-Ray intensity azimuthal profile from the 2D detector is not a direct representation of the geometric orientation distribution function. Such profile does not accurately take into account, for example, misalignment from scattering elements in the plane defined by the fibre and beam axis, i.e. rotational symmetry about the fibre axis is lost. Nevertheless, as indicated by Hermans et. al. in their seminal paper on cellulose [12] and pointed out by Northolt in his study on CF [13], the difference between the geometrical ODF and an WAXS reflection in these fibres is modulated by a factor of *cos(Θ)*, where *Θ* is the Bragg angle, and the difference is thus negligible for the graphite (002) reflection or the correlated SAXS reflection used in this work. Indeed, in micromechanical models for various polymer fibres, including fibres of poly(p-phenylene terephthalamide), a rigid-rod polymer, assuming a chain orientation distribution equivalent to a the azimuthal profile of a single (i.e. (200)) reflection also led to a successful description of the relation between compliance and chain alignment [14]. Corrections for azimuthal broadening have been proposed [7, 15], but their application to CNT fibres, a system far from the notion of a single graphite crystal and with such weak WAXS scattering, is far from trivial. Graphitic layers in CNTs and their bundles have no crystallographic registry between layers, as a consequence, scattering intensity for (hk0) reflections is very low (see radial profile in Figure 2) and the number of available reflections to the determine the ODF often limited to the (002).

*In-situ* Raman experiments were carried out using a Renishaw inVia microRaman spectrometer. All the measurements were performed with a laser of λ=533 nm wavelength for 10s exposure, with the fibre parallel to the polarization direction. The magnification used was ×20 which was equivalent to a spot size of around 1.6 µm. Stress transfer is analysed through the normalized change full width at half maximum of the 2D Raman band (ΔFWHM*), defined as

$$\Delta FWHM* = \frac{FWHM(\sigma) - FWHM_{initial}}{FWHM_{final} - FWHM_{initial}} \qquad (4)$$

Scanning Electron microscopy (SEM) was carried out with an FIB-FEGSEM Helios NanoLab 600i (FEI) at 10 kV. TEM images were taken using a Talos F200X FEI operating at 20 kV.

The present mechanical study was conducted on few-layer MWCNTs fibres, as reported in a previous work [5]. Their tensile properties were determined from a minimum of 10 conventional uniaxial tensile tests per sample type. In-situ synchrotron WAXS, SAXS and Raman measurements were conducted on 6 CNT fibre specimens.

## 3. Results

This study is performed on individual CNT fibre filaments of 20 - 30 µm in diameter (Fig. 1a). Figure 1b depicts the prototypical tensile response of this type of CNT fibres. The linear elastic regime extends up to 1% strain. After linearity is lost, the fibres enter the region of irreversible deformation (plastic regime)

[16]. This region is singular for two reasons: its long range, covering from 1% to almost 10% strain; and by the steady hardening, described by the fact that stress doubles from the yield stress to the sudden fibre failure.

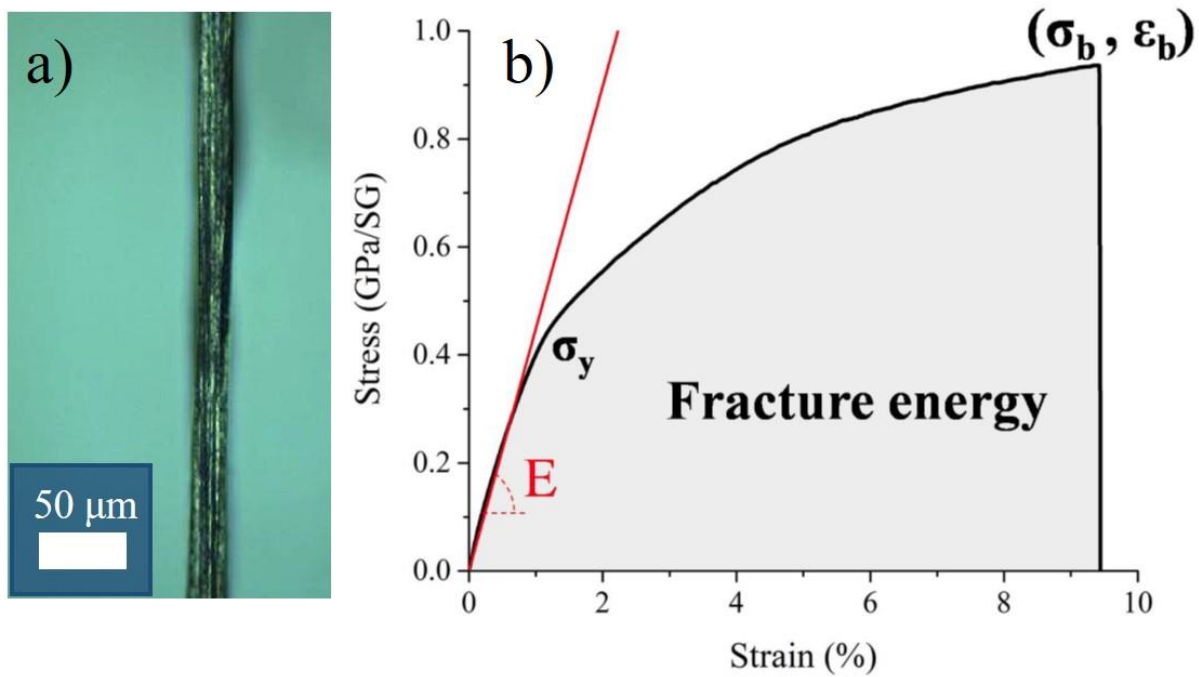

**Figure 1:** a) Optical micrograph of a macroscopic fibre of CNTs produced by the direct spinning method. b) Typical stress-strain curve showing unusual features for a high-performance fibre: large elastic-plastic transition region, hardening and large fracture energy.

No softening branch is observed. The combined effect of delayed failure and high tensile strength is the large fracture energy displayed by the fibres. The mechanisms responsible for this behaviour are discussed in this work.

Our strategy to understand the microstructural deformation processess is to perform *in-situ* WAXS/SAXS and Raman during tensile testing of CNT fibres. The starting point is to analyse the WAXS/SAXS radial profiles. In spite of the exceptional molecular perfection of the CNTs, they are very imperfectly packed into a turbostratic structure, manifested as a weak interlayer stacking (002) peak with an asymmetric lineshape extending to the SAXS region (Figure 2a). As expected, throughout tensile deformation reflection arcs on the equator get progressively narrower around the azimuthal angle, as shown in in the examples of azimuthal profiles from radial integration of the (002) as inset in Figure 2b. However, there are no appreciable changes in the radial position or intensity of the (002) peak radial profile. The main change observed is an intensity increase in scattering intensity in the region between the (002) peak and the form factor (FF) in SAXS, which can be clearly seen in the 2D patterns (Figure 2b). This is a consequence of the progressive alignment and aggregation of bundles upon stretching, leading to the pores adopting a more elongated shape and the emergence of the form factor along the equator. From SAXS data we calculate the changes in fibre porosity using the methodology reported earlier [11] (see experimental methods). Figure 2c shows that there is rapid reduction in porosity from 0.74 to 0.69 during tensile deformation of the fibre, which then levels off and remains fairly constant throughout the rest of the test. Note though, that this SAXS-derived porosity is with respect to Bernal graphite and include, for example, the inside of the CNTs For reference, assuming a theoretical density of 1.8*g/cm³* for a perfect CNT fibre, the change in CNT fibre density is from 0.47 to 0.56*g/cm³*.

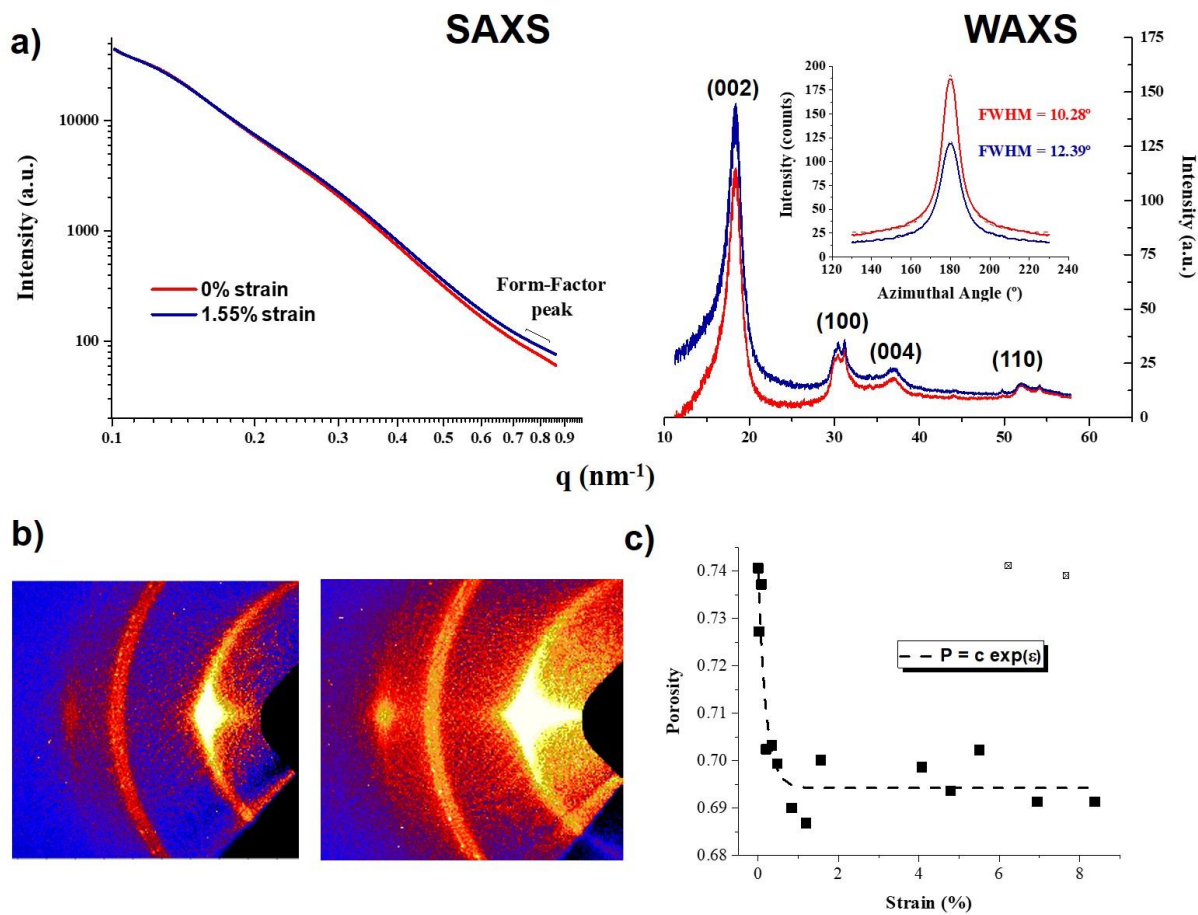

**Figure 2:** In-situ WAXS/SAXS during CNT fibre stretching. a) WAXS/SAXS radial profiles showing structural changes mainly in small pores, attributed to bundles packing closer together (inset: azimuthal profiles from radial integration of the (002) reflection). CNT (hkl) reflections marked. b) Examples of 2D WAXS patterns showing increased alignment and scattering intensity upon stretching. c) Decrease in porosity during tensile deformation determined from SAXS.

The picture that emerges from the analysis of radial profiles is that during stretching structural changes are dominated by densification as bundles pack closer together, but without the formation of new bundles or "crystallisation" of CNT through the formation of new turbostratic domains. This result is important, because it indicates that during the tensile deformation of the fibre there is no significant formation of new CNT overlap domains capable of stress transfer, and thus, that tensile deformation could be described exclusively through changes in bundle orientation.

Statistical rotational symmetry with respect to the central fibre axis allows the description of bundle orientation in terms of the polar angle only. Therefore, both the discrete polar angle histogram or its continuous counterpart the orientation distribution function (ODF) either defined in the [-π/2, π/2] or [-π, π] intervals, fully account for the orientation information. From an experimental viewpoint, such information is in principle accessible through direct image analysis, as from Fig. 1a, but this method can only provide orientation from the fibre surface and requires many images for the samples to be large enough to produce representative results [17]. Polarized Raman spectroscopy has also been successfully used to determine the ODF of CNTs *on the surface* of CNT fibres [18, 19, 20].

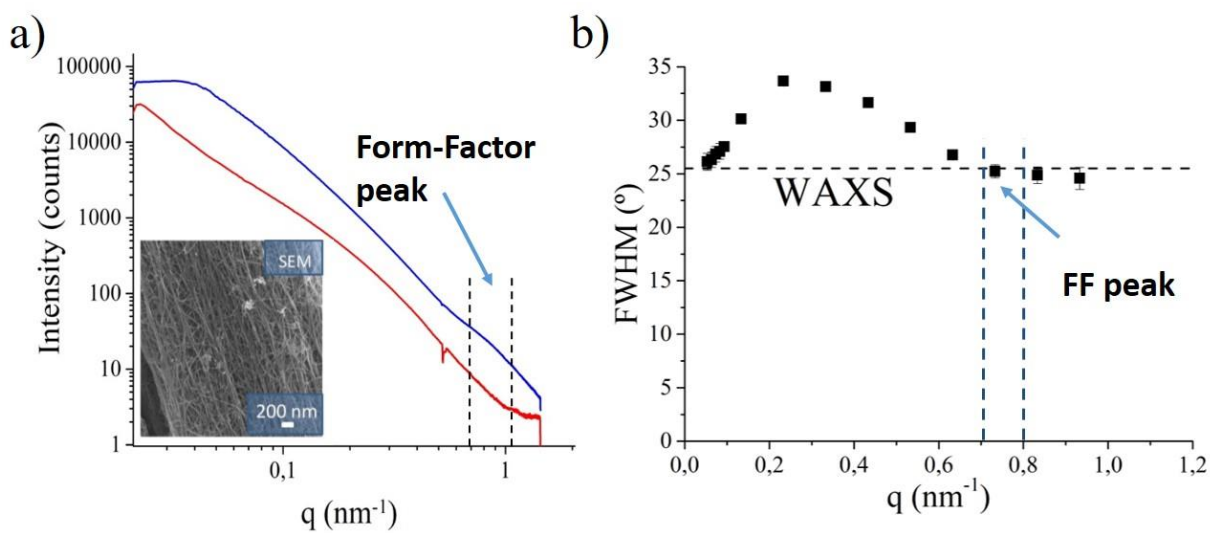

**Figure 3:** Determination of the ODF of load-bearing elements from synchrotron WAXS/SAXS. a) SEM micrograph of a CNT fibre showing preferential alignment of bundles, and radial profiles from azimuthal integration of SAXS patterns, showing the CNT form factor in the equatorial profile (blue - equatorial, red - meridional). b) The FWHM of the azimuthal distribution from WAXS (dashed line) matches the FWHM from SAXS (points) in the FF q region, enabling the use of more-intense SAXS to describe the ODF of load-bearing elements in the fibre.

X-ray scattering provides information from the whole irradiated volume and allows cross-validation between WAXS and SAXS [21]. WAXS is the technique of reference for the determination of orientation in highly crystalline fibres, including ordinary carbon fibres; however, WAXS intensity is very low in CNT fibres due to their intrinsically turbostratic structure, making SAXS more suitable to determine the ODF of *individual* CNT fibres. But since SAXS involves a wide range of scattering lengths, the first question is to determine the correct range of $q$ for azimuthal integration of SAXS. Through measurements on multi-filament samples we find that in the FF region, a small hump in the $q$ range 0.7-0.8 nm$^{-1}$ (Fig. 3a), there is excellent agreement in terms of the FWHM of both distributions for the reference material (25.5° in WAXS vs. 25.2° in SAXS, Fig. 3b). We thus use this q-range for analysis of the ODF ($\psi(\varphi)$) of individual CNT fibre filaments.

Equipped with the ODF from SAXS, it is then possible to analyse the changes in CNT throughout the tensile test. It was previously shown that fibre stiffness could be directly related to the initial orientation parameter <cos²(φ₀)> through the bundle longitudinal $e_c$ and shear $g$ moduli,

$$\frac{1}{E} = \frac{1}{e_c} + \frac{<\cos^2(\phi_0)>}{g} \tag{5}$$

This expression provides an accurate estimate of stiffness for fibres with a high degree of alignment [5], but fails to capture the mechanical response as deformation evolves. It is clear that any additional macroscopic stretch of the fibre will increasingly align the fibre bundles towards the load axis, leading to a narrower ODF and, consequently, to a lower orientation parameter. Rotation is accommodated in the bundle through shear deformation. By considering small, elastic strain increments, Northolt [14] used the exponential map and its first-order series expansion to express such alignment:

$$\frac{<\cos^2(\phi_0)>}{<\cos^2(\phi)>} = exp\left(\frac{\sigma}{g}\right) \approx 1 + \frac{\sigma}{g} \tag{6}$$

By rearranging terms and introducing the new variable U ≡ <cos²(φ₀)>/ <cos²(φ)> -1, we recover a linear relationship between stress and U:

$$\sigma = gU \tag{7}$$

U is a key dimensionless variable, equal to zero at the initial orientation, which behaves like an alignment strain (Figure 4a).

Fig. 4b shows the experimental stress-alignment (σ vs U) curve of CNT fibres produced at different winding rates. During the initial stage, eq.7 holds and the behaviour is linear. But from around 0.5% strain, the behaviour deviates from the expected response of a fully oriented ideal material, which would abruptly transition from elastic to perfectly plastic deformation and for which stress would increase linearly up to the yield stress and then remain constant as it undergoes a large reorientation. In contrast, the real material yields at around one half of the ultimate tensile stress. After yielding, we observe a decrease in the slope of the curve (loss of stiffness), but also an increment in stress. Orientation plays a fundamental role for both phenomena: on the one hand, individual bundles yield depending on their orientation, and therefore the elastic to plastic transition ceases to be sharp. On the other hand, additional alignment implies more bundles recruited along the fibre axis, and a net increase in the load carried by the fibre. In other words, these results show that the CNT fibre undergoes hardening upon progressive alignment of bundles. Both kinds of behaviour are schematically shown in Figure 4c. By cooperative behavior we denote the ability of interconnected structural elements to contribute to the mechanical properties of the ensemble. It is clear that bundles in CNT fibres display a cooperative behavior, but a variable degree of cooperativity can be observed depending on connectivity and orientation. In a random fibre network subjected to a 1D tensile load, some fibres might not be linked to the rest, might as well be perpendicularly oriented with respect to the load axis, or exhibit waviness. In all these cases, these fibres would not contribute to the load-bearing capacity of the network, which is left to an `active skeleton'.

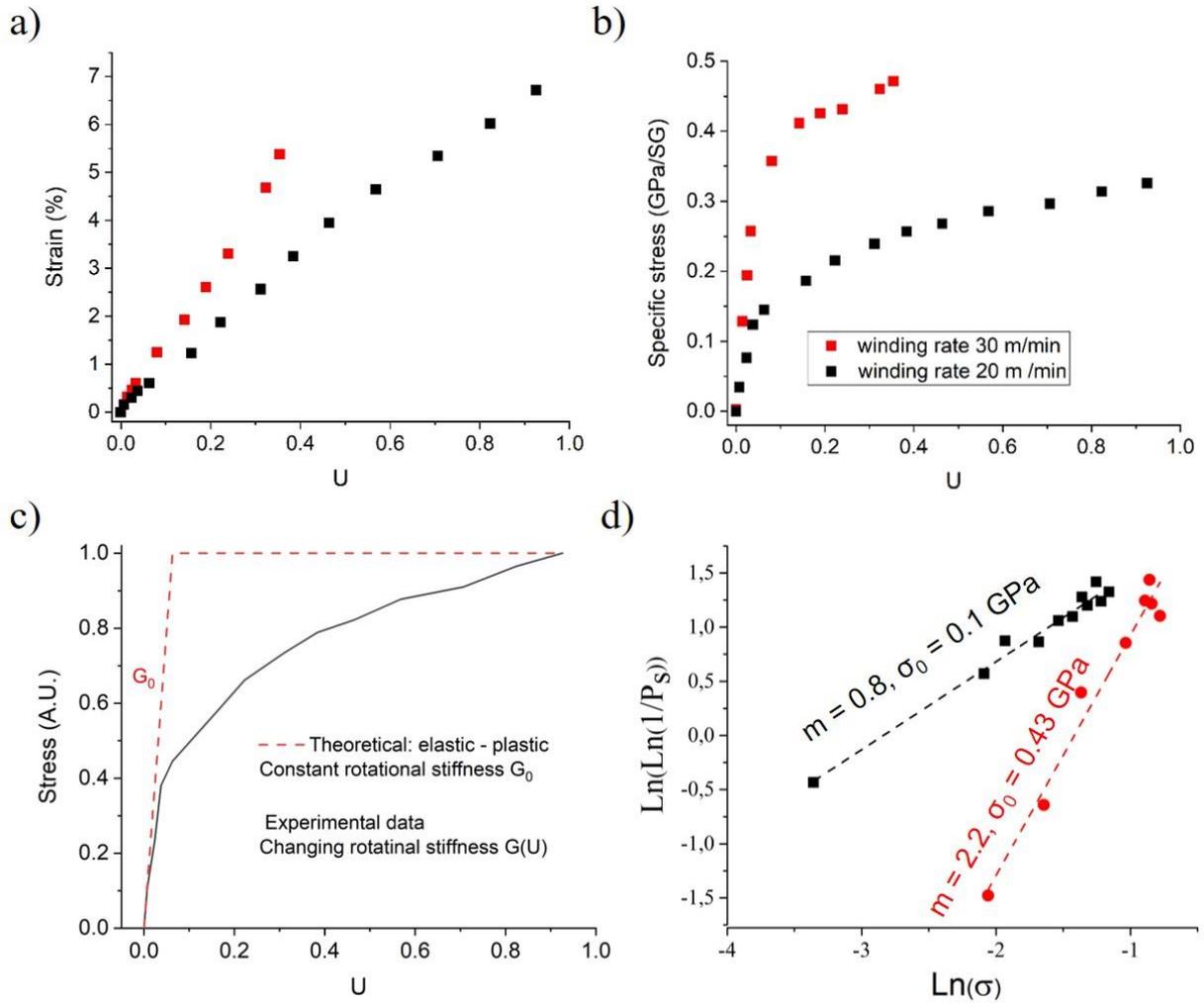

**Figure 4:** Evolution of CNT alignment during tensile testing as determined from SAXS. Plots of fibre strain a) and b) specific stress against the orientational parameter *U* showing that the changes in alignment capture the elasto-plastic lineshape of the stress strain curve, and the differences in degree of alignment resulting from winding (drawing) fibres at different rates. c) The changes in alignment and tensile stress during the tensile test resulting in a non-constant rotational stiffness, with d) a dependence following a Weibull distribution, interpreted as indication of progressive sliding of elements.

Rotational stiffness, defined as the slope of the curve $G = \Delta\sigma/\Delta U$, decreases in an almost monotonic way. This behaviour can be explained if bundle sliding is triggered within the fibre at different stress levels, which is fully compatible with the fact that the fibre was originally synthesised from an aerogel as a random network where bundles are arranged in many ways with different connectivity. The hypothesis of bundle sliding under a distribution of stress levels is confirmed by postulating a Weibull distribution and fitting to it the experimental data. In order to do so, the normalised rotational stiffness $G/G_0$, can be taken as a measurement of yielding. Clearly, $G/G_0 = 1$ implies that the fibre fully remains in the elastic regime without bundles having undergone any sliding, while the loss of rotational stiffness is the result of the failure of internal links and bundle sliding. Following the classical terminology of weakest-link theory, $P_f$ denotes the failure probability – here understood as bundle yielding through sliding-, while $P_s = 1 - P_f = G/G_0$ is the survival (not sliding) probability. Figure 3d shows that the experimental data can be fitted very accurately to a survival Weibull distribution

$$P_s = exp\left[-\left(\frac{\sigma}{\sigma_0}\right)^m\right], \tag{8}$$

where $\sigma_0$ is a stress scale factor and $m$ the shape factor, or Weibull modulus of the distribution. The scale factor $\sigma_0$ has a probabilistic meaning, since it is the stress level at which 63% of elements have yielded. We briefly recall that a large value for the Weibull modulus ($m \to \infty$) would imply a sharp, fully deterministic value for the yield stress at , recovering the elastic, perfectly plastic response sketched in Fig. 4 (c). Lower values of $m$ broaden the distribution and sliding takes place over an increased range of stress. To determine the values of $m$ for both fibres, a Weibull plot is employed (Fig. 4d), where $m$ is obtained as the slope of the double logarithm of $P_s$ vs. the logarithm of stress.

From Figure 4d is also possible to compare Weibull parameters for CNT fibres produced at 20m/min and 30m/min. Clearly, the values of m prove that sliding of bundles for more aligned fibres (30 m/min) occurs at a narrower stress interval compared to fibres with the same chemical composition but lower degree of alignment (20 m/min). This fact indicates that in more oriented fibres bundles need higher stresses to slide, but they do so in a more cooperative way, displaying lower deviation from central values.

Raman spectroscopy provides valuable information in order to complete this analysis. Since the position and shape of the 2D is linked to the stretching of $sp^2$-$sp^2$ bonds in CNTs [22], the Raman signal can be used as a bundle stress probe. On the one hand, the peak linearly shifts towards lower wavenumbers upon tensile loading. Fig. 5(a) shows the 2D spectrum when the macroscopic fibre is unloaded (red curve) and after reaching 4% strain (black). Two effects are noticed: the peak position is shifted to the left (lower wavenumbers) and an asymmetric broadening, also to the left, occurs. The figure inset displays a plot of the peak position vs. fibre strain. As can be observed, there is a linear dependency of the shift with respect to strain in the initial stage of deformation. Beyond this point (near 0.5% strain), no additional shift is detected. This saturation indicates that a sliding mechanism is triggered and the bundles cease to carry additional load. The stress level at which sliding occurs is a probabilistic variable: for each bundle, it depends on the bundle connectivity and orientation, but globally, its statistical parameters reveal how cooperative the network is.

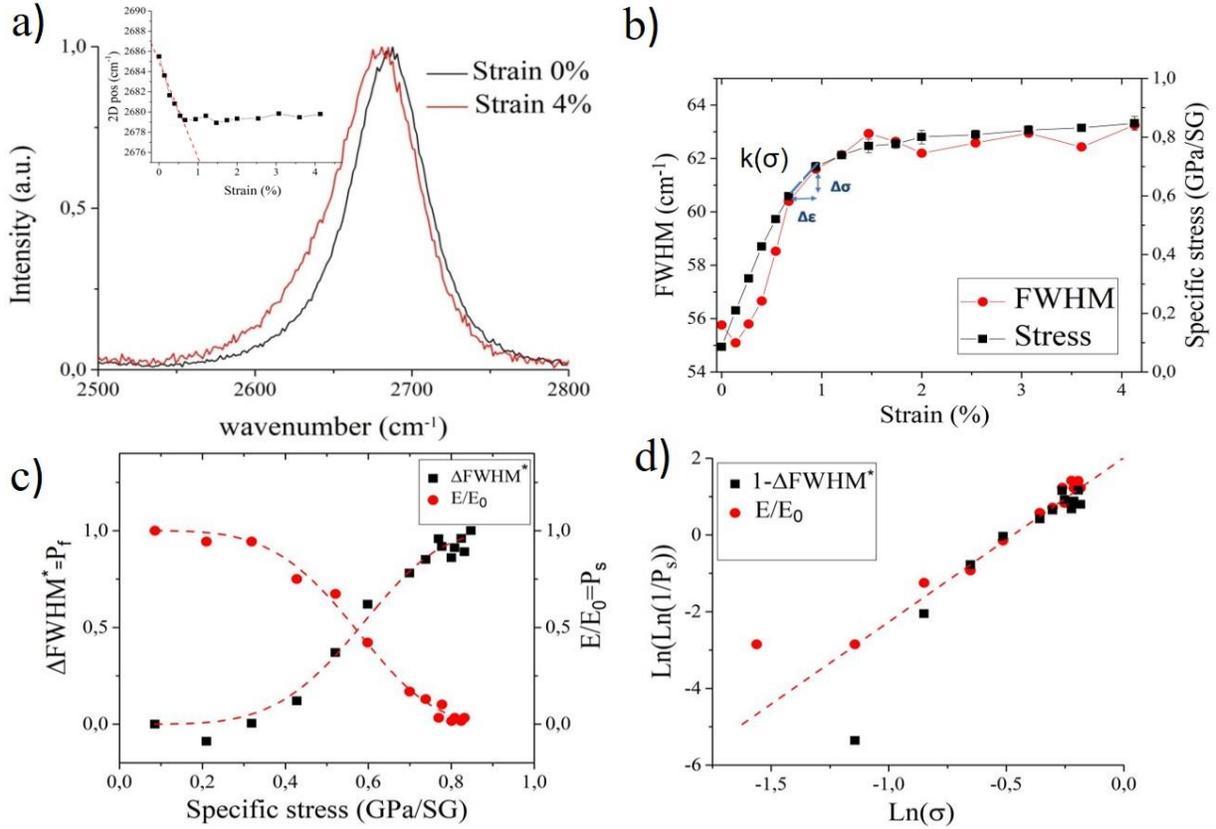

**Figure 5:** In-situ Raman spectra during tensile testing. a) Raman downshift (inset) and asymmetric peak broadening. b) Progressive FWHM increase with increasing bulk tensile stress. c) and d) Both instant modulus and FWHM follow a Weibull distribution with applied stress.

It is instructive to analyse the peak broadening upon stretching, observed generally in high-performance fibres [23] and similar CNT fibres [16, 24], and taken to reflect a distribution of stress states. Figure 5 (b) presents a direct comparison of the peak FWHM vs. strain and stress-strain curve. As can be observed, both curves present the same behaviour over the whole strain domain. Now, the analysis is extended by adopting a similar strategy to the stress-orientation case. Instead of rotational stiffness, we take now into consideration the usual instantaneous (tangent) stiffness, $k = Δσ/Δε$. Its normalized version, $k/k_0$, can be taken as our primary variable. If we plot both this variable and the normalised FWHM increment (ΔFWHM*) vs. specific stress (Fig. 5c), it is immediately noticed that both curves are complementary (i.e. their sum equals 1). Furthermore, both variables can again be described with Weibull distributions: the normalised tangent stiffness by the survival Weibull distribution (Eq.8), and the extra Raman broadening -FWHM- by its failure complementary distribution:

$$P_f = 1 - exp\left[-\left(\frac{\sigma}{\sigma_0}\right)^m\right], \quad (9)$$

Figure 5(d) depicts the excellent match of both variables in a Weibull plot, which provides a Weibull modulus value for the normalised tangent stiffness $m$ =4.3.

A similar exponential dependence of longitudinal modulus on axial stress has been recently proposed for a range of CNT fibres also produced by floating catalyst chemical vapour deposition. [25]

Effectively, these results show that CNT fibres behave as a weakly cooperative network. Upon loading there is progressive sliding of CNT bundles, directly observed through changes in alignment monitored by SAXS, which corresponds to a non-uniform stress distribution in the CNTs, probed by Raman spectroscopy. As such, the Weibull modulus becomes a key descriptor of the material. Thus, we find that the sample produced at higher draw ratio is not only more aligned, but has a higher value of m, indicating more cooperative loading. This correspondence also helps explain the widespread observation that CNT fibre strength and modulus are proportional [26].

The degree of cooperative loading, embedded in $m$, is particularly sensitive to the initial CNT fibre structure. In-situ Raman measurements during loading/unloading cycles, included in Figure 6, show a tail in the initial stress-strain curve, typically associated with CNT waviness. Correspondingly, $m$ = 1.06. During the second loading cycle m increases to 4.53; load is shared more cooperatively after alignment induced in the preceding cycle.

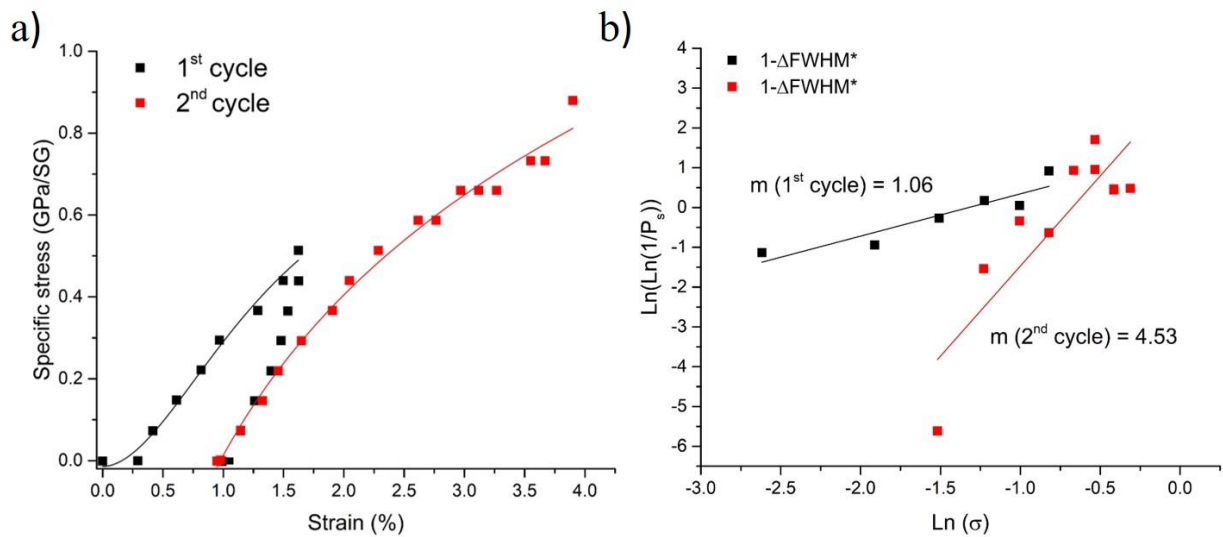

**Figure 6:** In-situ Raman spectra during load-unload tensile testing. a) Stress-strain curve showing a slight increase in modulus and yield stress. b) The 2D band FWHM increases according to a Weibull distribution, with modulus going from 1.06 to 4.53 upon re-loading due to a more even stress transfer after bundle slippage in the first cycle.

The fact that Weibull modulus from alignment measurements (i.e. SAXS) is lower that the Raman/tangent stiffness implies that the progressive realignment of bundles and load distribution are not fully correspondent, since they are formulated with respect to two different variables (stiffness and shear stiffness, respectively) and mechanisms. For instance, differences in CNT morphology with respect to diameter and number of CNTs layers have strong effect both on resonant Raman scattering intensity and on load-bearing capacity. But in spite of the current limitation in reducing Weibull parameters to simple fibre features, both sets of parameters are meaningful and can be conveniently used to compare fibres or fibre families. We also note that the Weibull modulus obtained from tensile tests on traditional staple fibres is a similar range (1.26 – 1.77) [27], supporting the view that CNT fibres behave as yarns made up of nanoscale subfilaments.

Finally, there other parameters that can be extracted from the Weibull analysis using SAXS data. The shear rotational stiffness $G_0$, extracted from the linear part of Figure 4a comes out as 5.6 – 16 GPa. This parameter has close correspondence to the shear modulus of the bundles and is in the range observed for carbon fibres 5 – 35 GPa [28, 29]. Similarly, the yield point in the stress *vs. U* plot can be related to a shear strength by

$$\tau = \sigma_y < |sin\phi_y cos\phi_y| > \qquad (10)$$

The experimental values obtained were between 22.4 and 40.8 MPa, which are in the range of values reported from tests on individual CNTs/bundles with sufficiently long CNT overlap length for stress to saturate under shear-lag based stress transfer (30 – 69 MPa) [30, 31].

## Conclusions

The micromechanics of CNT fibres during tensile deformation were studied by performing in-situ structural characterisation. Using in-situ synchrotron WAXS/SAXS we could monitor the changes in degree of alignment during stretching through the evolution of the orientation distribution. The data show an initial region of small changes in alignment per applied tensile stress, where bulk strain is dominated by elastic deformation of bundles and their rotation, followed by extensive bundle sliding and alignment. This behaviour can be described by a Weibull distribution. Similarly, in-situ Raman spectroscopy is used to monitor the stress in the CNTs through changes in the FWHM of the 2D band. The evolution of FHWM with applied stress also follows a Weibull distribution, confirming the re-distribution of stress in the bundles upon sliding. The tangent stiffness follows a similar behaviour and gives the same Weibull modulus as the Raman plot. The Weibull modulus becomes an important descriptor of cooperative loading in the samples, and it is thus expected that CNT fibres with higher tensile strength produced by other spinning methods have higher values of *m*. However, the exceptionally high fracture energy of the samples produced from floating catalyst CVD is likely related to their moderate values of *m*, which enables large energy absorption through sliding of the extraordinarily long CNTs. The prospect is that independent control of the ODF and Weibull modulus m could enable tailoring of tensile properties. This requires translating the parameter m into a descriptor of the material. This view is in agreement with the statistical treatment of hierarchical composites provided by Wu et al. [32]. Performing similar *in-situ* studies using direct imaging techniques, such as electron microscopy [33], could help identify structural features of the CNT fibre network related to the parameters extracted from the Weibull analysis.

Additional parameters that emerge from this analysis are an apparent bundle shear modulus, with values of 5.6 – 16 GPa, and a bundle shear strength of 22.4 and 40.8 MPa. Finally, from analysis of WAXS and SAXS we can confirm that during stretching, the large increase in degree of alignment produces narrowing of the pores between bundles and an overall increase in density, but without indication of induced crystallization through the formation of CNT-CNT domains sufficiently close for stress transfer.

## Acknowledgements

The authors acknowledge generous financial support from the European Union's Seventh Framework Program under grant agreement 678565 (ERCSTEM) and Horizon 2020 research and innovation programme under the Marie Skodowska-Curie grant agreement 797176 (ENERYARN), and from MINECO (RyC-2014-15115). The SAXS-WAXS experiments were performed at NCD-SWEET beamline at ALBA synchrotron with the collaboration of ALBA staff.